\documentclass[aps,pra,twocolumn,amsmath,amssymb,bibnotes]{revtex4}

\usepackage{color,epsfig,rotating,graphicx,subfigure}

\usepackage[latin1]{inputenc}
\usepackage[english]{babel}

\usepackage{dsfont}
\usepackage{textcomp}

\renewcommand{\Im}{{\rm Im}}

\newcommand{\ri}{{\rm i}}
\newcommand{\re}{{\rm e}}
\newcommand{\rd}{{\rm d}}

\newcommand{\kb}{k_{\rm B}}

\begin{document}
\title{Radiative heat flux through a topological Su-Schrieffer-Heeger chain of plasmonic nanoparticles}

\date{\today}

\author{Annika Ott and Svend-Age Biehs$^{*}$}
\affiliation{Institut f{\"u}r Physik, Carl von Ossietzky Universit{\"a}t, D-26111 Oldenburg, Germany}
\email{ s.age.biehs@uni-oldenburg.de} 


\begin{abstract}
We investigate the radiative heat transport along a Su-Schrieffer-Heeger chain of InSb nanoparticles. We show that in the topological non-trivial phase, the edge modes dominate the radiative heat transport despite their strong localization at the edges of the finite chain due to a long-range coupling of the first and last particle. We further discuss the scaling laws of the heat transfer with respect to the chain length for the longitudinal and transversal band- and edge modes and conclude that both type of modes obey the same scaling law.
\end{abstract}

\maketitle


\section{Introduction}

The advent of many-body theories for describing nanoscale radiative heat transfer
in plasmonic and phonon-polaritonic systems on the basis of fluctuational electrodynamics~\cite{PBAmanybody,KruegerEtAl2012,nteilchen,DongEtAl2017,LatellaEtAl2017} opened up the way for a plethora of studies ranging from the investigation of ballistic, diffusive, sub- and superdiffusive transport regimes in many-body systems~\cite{PBAEtAl2013,LatellaEtAl2018,Kathmann,Tervo,Tervo2}, to investigations of dynamical control of heat fluxes, thermal relaxation dynamics, transistors and logic gates working with thermal radiation~\cite{Drag,Add4,Add5,PBASAB2014,OrdonezEtAl2016,KathmannEtAl2020,Carlos,RMP} and the application of the many-body theory in thermal discrete dipole approximation~\cite{Edalatpour2014,EkerothEtAl2017}. Recently, in particular the impact of non-reciprocity~\cite{moncadavilla,Song,WuEtAl,Herz}  in many-body systems has triggered the discovery of astonishing effects like persistent heat currents and heat fluxes~\cite{zhufan,zhufan2,Silveirinha,meinpaper}, persistent spins and angular momenta of the thermal near- and far field~\cite{meinpaper,Zubin2019}, giant magneto-resistance~\cite{Latella2017, Cuevas}, the Hall effect for thermal radiation for magneto-optical materials~\cite{hall,OttEtAl2019}, and the dynamical control via magneto-optical surface waves~\cite{paper_diode}. Finally, it has been recently shown that there is a radiative anomalous Hall effect in topological Weyl semi-metals~\cite{OttEtAl2020}, whereas in general the near-field radiative heat flux in topological many-body systems remains completely unexplored.

\begin{figure}[h!]
	\includegraphics[width=0.45\textwidth]{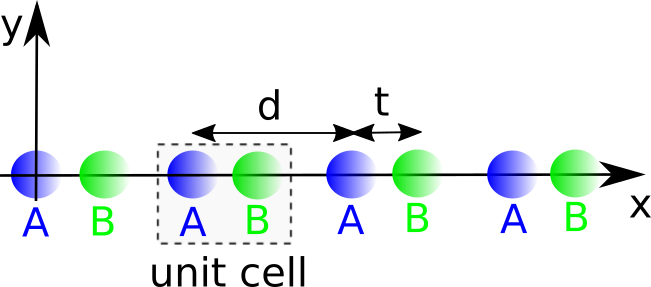}
	\caption{Sketch of the bipartite SSH chain of spherical NPs along the x axis with lattice constant $d$ and $t = \beta d/2$.}
	\label{fig:SSHchainSketch}
\end{figure}

In this work, we make the first steps in studying such toplogical systems by investigating the radiative heat transfer through 
a Su-Schrieffer-Heeger (SSH) chain made of plasmonic InSb nanoparticles (NP) as sketched in Fig.~\ref{fig:SSHchainSketch}. In such plasmonic chains, the transversal and longitudinal modes each form typically two bands of band modes separated by a bandgap which depends on the parameter $\beta$. 
As typical for the SSH model, in the topological non-trivial phase where $\beta > 1$ two topological protected edge modes 
appears in the bandgap when chains of finite length are considered~\cite{OESSH,ACSphotonSSH,JAPSSH} which are due to the 
symmetric and anti-symmetric coupling of the edge modes on the A and B sublattices. Naively, one might expect that these edge modes do not contribute to the radiative heat transfer along very long SSH chain, but that the radiative heat transfer in this case 
is largely due to the band modes. This expectation can be made by analogy with the radiative heat transfer between two semi-infinite materials. In this situation the radiative heat flux is dominated by propagating waves when the distance between semi-infinite materials is larger than the thermal wavelength $\lambda_{\rm th}$ which is approximately $10\,\mu{\rm m}$ at room temperature. Now, if the semi-infinite materials support surface waves in the infrared then there will be a coupling of these surface wave when the interfaces are approached to distances much smaller than $\lambda_{\rm th}$. This coupling via the evanscent fields of the surface waves provides a new heat flux channel which can lead to near-field heat fluxes which are orders of magnitude larger than the blackbody value as predicted theoretically and demonstrated experimentally~\cite{Carlos,RMP,Fiorino2018}. By this analogy, one might expect that the edge modes on the A and B sublattice of the SSH chain can couple and lead to a large heat flux when the chain length is small. On the other hand, for very long SSH chains this evanescent coupling of the edge modes should become weak and therefore the heat flux might be due to the propagating
modes which are the band modes. Here, we demonstrate that contrary to this naive expectation the edge modes dominate 
the radiative heat transfer along the SSH chain for chains with arbitrary length so that even for SSH chains which are much longer than the thermal wavelength $\lambda_{\rm th}$ the heat flux by the edge modes is more important than that by the band modes. Our work is organized as follows. In Sec.~II we review the mode structure and the topological trivial and non-trivial phases of a SSH NP chain. In Sec.~III we provide the fundamental relations for the radiative heat flux and the material properties of the NPs needed for the numerical evaluations. The numerical results are discussed in Sec.~IV and the general conclusions are given in Sec. V.

\section{Mode structure and Zak phase}

We consider a bipartite chain of identical spherical isotropic NPs with a lattice constant $d$ and the spacing
between NPs A and B of a unit cell of $t = \beta d/2$ as depicted in Fig.~\ref{fig:SSHchainSketch}. 
The induced polarization $\mathbf{p}_{Ai}$ and $\mathbf{p}_{Bi}$ of the NPs A an B in the i-th 
unit cell at frequency $\omega$ due to all other NPs can be expressed as
\begin{equation}
	\mathbf{p}_i = k_0^2 \sum_{j \neq n} \alpha \mathds{G}(x_i,x_j) \mathbf{p}_j
	\label{Eq:Eigenmodes}
\end{equation}
where $i$ and $j$ run over all NPs $Ai$, $Bi$ and $Aj$, $Bj$; $k_0 = \omega/c$ is the vacuum wavenumber and $c$ is the light velocity in vacuum. The vacuum Green function~\cite{Novotny}
\begin{equation}
\begin{split}
	\mathds{G}(x_i,x_j) &= G_\parallel(x_i - x_j) \mathbf{e}_x \otimes \mathbf{e}_x \\
	                    &\quad +  G_\perp(x_i - x_j) \bigl[ \mathbf{e}_y \otimes \mathbf{e}_y + \mathbf{e}_z \otimes \mathbf{e}_z\bigr]
\end{split}
\end{equation}
with
\begin{align}
	G_{\parallel}(x_i - x_j) &=  \frac{e^{i k_0 |x_i - x_j|}}{4\pi |x_i - x_j|} (a + b), \\
	G_{\perp}(x_i - x_j) &=  \frac{e^{i k_0 |x_i - x_j|}}{4\pi |x_i - x_j|} a  
\end{align}
and
\begin{align}
	a &= 1+\frac{ik_0|x_i - x_j|-1}{k_0^2|x_i - x_j|^2}, \\
	b &=\frac{3-3ik_0|x_i - x_j|-k_0^2|x_i - x_j|^2}{k_0^2|x_i - x_j|^2}.
\end{align}
expresses the dipole field at positions $x_i$ generated by the NP at position $x_j$ and $\alpha$ is the polarizability tensor of the NPs. 

As discussed in detail in Refs.~\cite{OESSH,ACSphotonSSH,JAPSSH}, for an infinite SSH chain by using the Bloch theorem this equation can be brought into the form of a simple eigenvalue equation
\begin{equation}
	\mathds{M}_\mathbf{\nu} \begin{pmatrix} p_{\nu, A} \\ p_{\nu, B} \end{pmatrix} = \frac{1}{\alpha}\begin{pmatrix} p_{\nu, A} \\ p_{\nu, B} \end{pmatrix}
\end{equation}
with
\begin{equation}
	\mathds{M}_\mathbf{\nu} = \begin{pmatrix} M_{\nu,AA} & M_{\nu,AB} \\ M_{\nu,BA} & M_{\nu, BB} \end{pmatrix}
\end{equation}
and
\begin{align}
	M_{\nu,AA} &=  M_{\nu,BB} = k_0^2  \sum_{j \in \mathds{Z}, j \neq 0} G_\nu (jd)\mathbf{e}_\nu \re^{\ri k_x j d},  \\
	M_{\nu,AB} &=  k_0^2 \sum_{j \in \mathds{Z}} G_\nu (jd + t) \re^{\ri k_x j d}, \\
	M_{\nu,BA} &=  k_0^2 \sum_{j \in \mathds{Z}} G_\nu (jd - t) \re^{\ri k_x j d} 
\end{align}
for the two polarizations perpendicular and parallel to the chain $\nu = \perp, \parallel$.

\begin{figure}[h!]
	\includegraphics[width=0.45\textwidth]{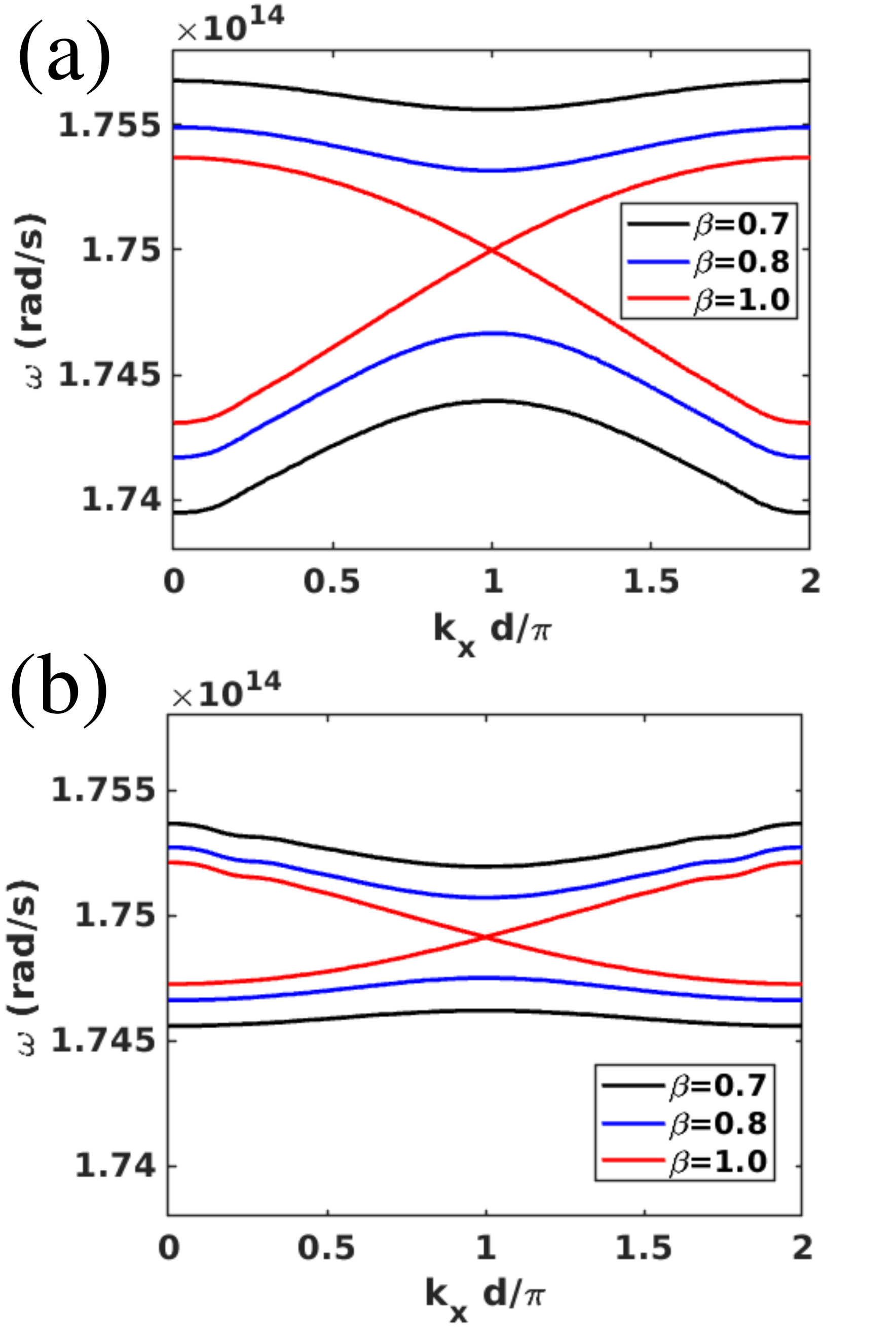}
	\caption{Longitudinal (a) and transversal (b) eigenmodes determined by $\det(\mathds{M}_\nu - \frac{1}{\alpha} \mathds{1}) = 0$ of an infinite chain of InSb nanoparticles with radius $R = 100\,{\rm nm}$ and lattice constant $d = 1\,\mu{\rm m}$ for $\beta = 0.7, 0.8$ and $1.0$.}
	\label{fig:Banddiagramm}
\end{figure}

The eigenmodes of the infinite SSH chain are determined by $\det(\mathds{M}_\nu - \frac{1}{\alpha} \mathds{1}) = 0$ which form typically for each polarization $\nu$ two bands as can be seen in Fig.~\ref{fig:Banddiagramm}. Due to the symmetry the bands with $\beta = 1 + x$ and $\beta = 1 - x$ for a given value $x$ are the same and for $\beta = 1$ both bands merge into a single band, i.e.\ the band gap is closed. In order to reveal the topological features of the SSH chain, one can focus on the quasi-static regime $k_0 d \ll 1$ (and then also $k_0 t \ll1$) where the quantities $G_\nu$ become real valued. Then the matrix $\mathds{M}_\nu$ can be expressed in terms of the Pauli matrices $\sigma_{x,y,z}$, as any two-band Hamiltonian, as ($\nu = \perp,\parallel$)
\begin{equation}
	\mathds{M}_\mathbf{\nu}^{\rm qs} = g_{\nu,0} \mathds{1} + \boldsymbol{\sigma} \cdot \boldsymbol{g}_\nu
\end{equation}
with $g_{\nu,0} = M_{\nu,AA}^{\rm qs} = M_{\nu,BB}^{\rm qs}$ and real valued $\mathbf{g}_\nu = (g_{\nu,x}, g_{\nu,y}, g_{\nu,z})^t$ with $g_{\nu,x} = (M_{\nu,BA}^{\rm qs} + M_{\nu,AB}^{\rm qs})/2$, $g_{\nu, y} = (M_{\nu,BA}^{\rm qs} - M_{\nu,AB}^{\rm qs})/2 \ri$, and $g_{\nu, z} = 0$. Obviously, $g_0$ breaks the chiral symmetry, but this is only a trivial symmetry breaking~\cite{ACSphotonSSH}. The chiral symmetry is reinforced for the matrix $\mathds{H}^{\rm qs} := \mathds{M}_\mathbf{\nu}^{\rm qs} - g_{\nu,0} \mathds{1}$, for which $\sigma_z \mathds{H}^{\rm qs} \sigma_z = - \mathds{H}^{\rm qs}$. This chiral symmetry property is still present in the full retarded regime. Hence, the bipartite NP chain is indeed a SSH chain and exhibits a topological phase transition with respect to $\beta$. The topological phase can be quantified by the Zak phase of one of the two bands for each polarization $\nu$ which can be expressed in our system by~\cite{ACSphotonSSH}
\begin{equation}
\begin{split}
	\gamma_\nu = \frac{\ri}{4} \log\biggl(\frac{M_{\nu,AB}^{\rm qs}}{M_{\nu,BA}^{\rm qs}}\biggr) \biggr|^{kx = \pi/d}_{k_x = - \pi/d}. 
\end{split}
\end{equation}
In the topological trivial case $\gamma_\nu = 0$ and in the topological non-trivial case $\gamma_\nu = \pi$. Typically, for a plasmonic NP SSH chain $\gamma_\nu = 0$ for $\beta \leq 1$ and $\gamma_\nu = \pi$ for $\beta > 1$ for both polarizations $\nu$ as shown in Refs.~\cite{OESSH,ACSphotonSSH,JAPSSH}. As a consequence, for a finite SSH NP chain there will be two topologically protected edge modes (for even $N$) in the band gap for $\beta > 1$ which do not exist for $\beta \leq 1$~\cite{OESSH,ACSphotonSSH,JAPSSH}.


\section{Radiative heat flux}

\subsection{heat flux formula}

We want to study the radiative heat flux through a finite SSH chain of $N$ NPs. In this case the topological properties of the infinite SSH chain persists, but the eigenmode frequencies and dipole moments~\cite{ford} need to be determined directly from Eq.~(\ref{Eq:Eigenmodes}). The general expression of the radiative mean power absorbed by NP $i$ at temperature $T_i$ due to the heat flow from all other NP $j$ at temperatures $T_j$ can be determined within the approach of fluctuational electrodynamics and reads~\cite{RMP}
\begin{equation}
	\begin{split}
		\mathcal{P}_{i}  &=  \sum_{j \neq i } \int_0^\infty \frac{\rd \omega}{2 \pi}\, \hbar \omega \bigl[n_j - n_i\bigr] \bigl( 2 \mathcal{T}_{j \rightarrow i,\perp } +  \mathcal{T}_{j \rightarrow i,\parallel }\bigr) \\
				 &= \int_0^\infty \frac{\rd \omega}{2 \pi}\, (2 \mathcal{P}_i^{\perp}(\omega) + \mathcal{P}_i^{\parallel}(\omega))
	\end{split}
\end{equation}
where $n_{i/j} = 1/(\exp(\hbar \omega/ \kb T_{i/j}) - 1)$ are the bosonic mean occupation numbers at temperatures $T_{i/j}$ and $\mathcal{T}_{j \rightarrow i,\nu}$ is the transmission coefficient from particle $j$ to $i$ for the polarization $\nu$. When omitting the negligible radiation correction it can be written as~\cite{RMP}
\begin{equation}
	\mathcal{T}_{j \rightarrow i, \nu} = 4 \frac{\Im(\alpha)^2}{|\alpha|^2}  \boldsymbol{T}_{\nu, ij}^{-1} \bigl(\boldsymbol{T}_\nu^{-1})^{\dagger}_{ij}  
	\label{Eq:TransmissionKoeff}
\end{equation}
introducing the $N\times N$ matrices
\begin{equation}
	\boldsymbol{T}_{\nu,ij} = \delta_{ij}  - (1 - \delta_{ij}) k_0^2 \alpha G_\nu (x_i - x_j).
\end{equation}
Note, that these expressions have been derived within the dipole approximation which is only valid as long as the radii $R$ of the NPs are much smaller than the wavelength so that $k_0 R \ll 1$ and $t > 4R$ and $d-t > 4R$ assuring that the distance between adjacent NPs is at least four times larger than the radii of the NP. For smaller distances also multipolar contributions will contribute to the radiative heat flux~\cite{Naraynaswamy2008,PBA2008,Otey,Becerril}.

\subsection{Material Properties} 

In the following we consider spherical InSb NPs with radius $R$ having a polarizability in the quasi-static limit ($R k_0 \ll 1$) given by
\begin{equation}
	\alpha = 4\pi R^3\frac{\epsilon-1}{\epsilon + 2}.
\end{equation}
Furthermore, we are using only the dominant electronic part of the optical response modelled by the Drude permittivity~\cite{exp}
\begin{equation}
   	\epsilon = \epsilon_\infty \left(1 - \frac{\omega_{\rm p}^2}{\omega(\omega+{\rm i}\Gamma)} \right),
\label{eps1}
\end{equation}
with the effective mass $m^* = 7.29\times10^{-32}$ kg, the density of the free charge carriers $n = 1.36\times10^{19}$ cm$^{-3}$, the dielectric constant for infinite frequencies $\epsilon_\infty = 15.68$ and the damping constant $\Gamma = 1\times10^{12}\,{\rm s}^{-1}$. The resonance frequency of the localized plasmonic modes in the InSb NP is with these parameters $\omega_{\rm LP} = \omega_p \sqrt{\epsilon_\infty/(\epsilon_\infty + 2)} = 1.752\times10^{14}\,{\rm rad}{\rm s}^{-1}$, i.e.\ it clearly lies in the infrared regime around $\lambda_{\rm th} \approx 10\,\mu{\rm m}$ which is relevant for thermal radiation around room temperature.

\begin{figure}[hbt]
	\includegraphics[width=0.45\textwidth]{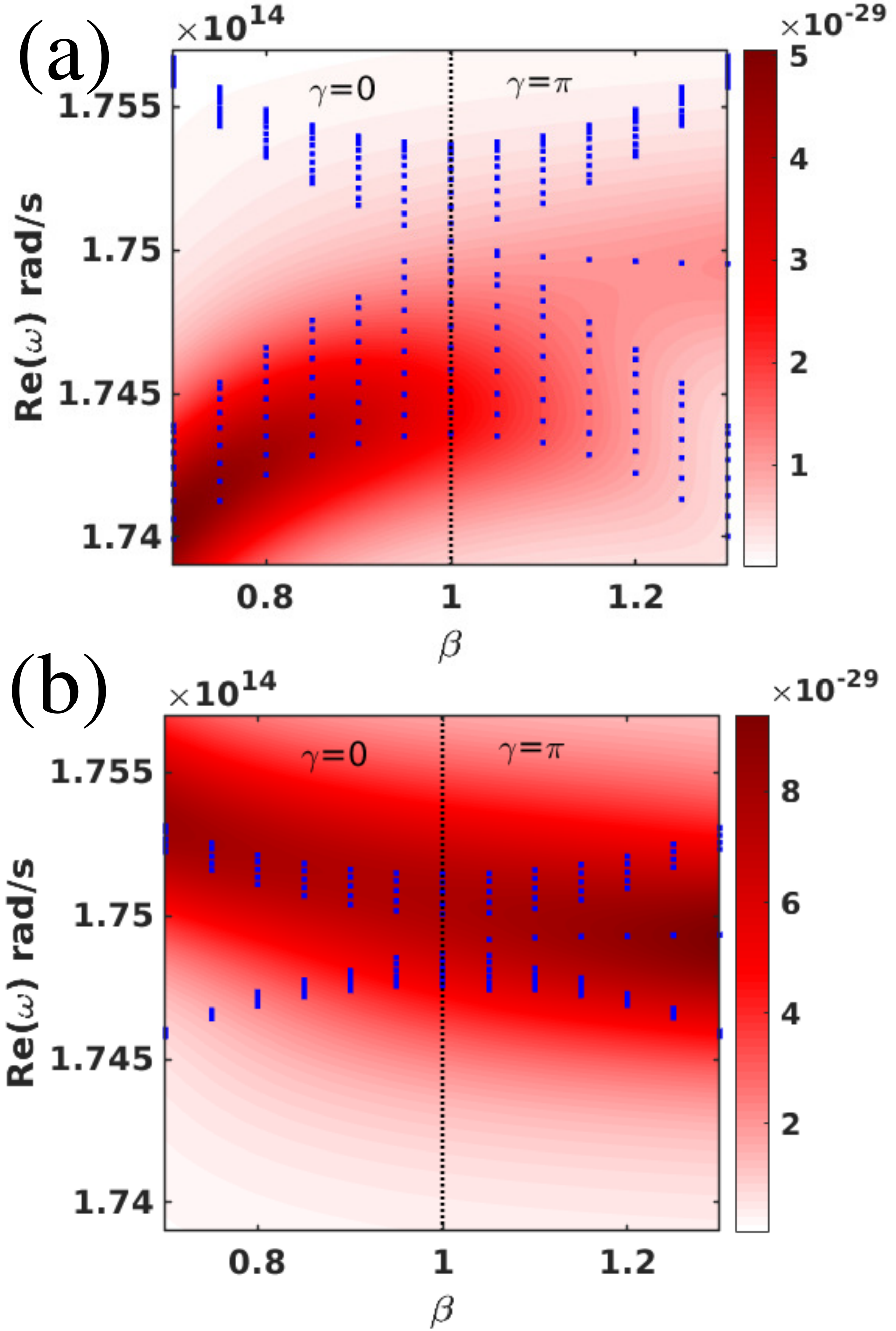}
	\caption{$\mathcal{P}_N^\nu (\omega)$ for the longitudinal modes ($\nu = \parallel$) in (a) and for the transversal modes ($\nu = \perp$) in (b) using $N = 20$ InSb NPs with radius $R = 100\,{\rm nm}$ and lattice constant $d = 1\,\mu{\rm m}$ where the temperature of the first NP is $T_1 = 310\,{\rm K}$ and all other NPs have temperatures $T_{j} = 300\,{\rm K}$ ($j = 2, \ldots, N$). The blue dots are the real parts of the complex eigenfrequencies of the chain modes determined by Eq.~(\ref{Eq:Eigenmodes}).}
	\label{fig:PNomegabeta}
\end{figure}

\begin{figure}[hbt]
	\includegraphics[width=0.45\textwidth]{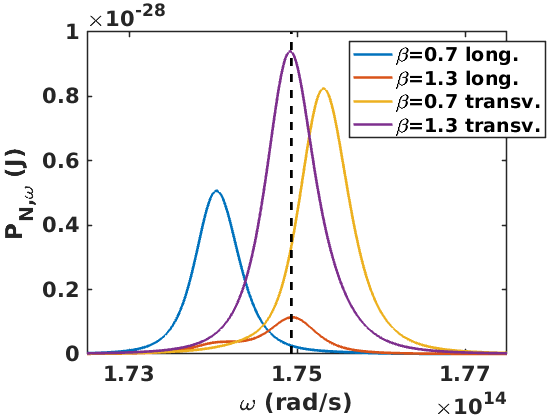}
	\caption{$\mathcal{P}_N(\omega)$ for the longitudinal modes ($\nu = \parallel$) and for the transversal modes ($\nu = \perp$) for $\beta = 0.7$ and $1.3$. The vertical dashed line is at the edge mode frequencies $\omega_{\rm em,\perp} \approx \omega_{\rm em,\parallel} \approx 1.493\times10^{14}\,{\rm rad/s}$ which practically coincide for the two transversal and longitudinal edge modes. We use the parameters are the same as in Fig.~\ref{fig:PNomegabeta}. The horizontal line devides the diagramm into the topological trivial region with Zak phase $\gamma_\nu = 0$ for $\beta \leq 1$ and the topological non-trivial region with Zak phase $\gamma_\nu = \pi$ for $\beta > 1$.}
	\label{fig:PNomega}
\end{figure}

\section{Numerical results}

In Fig.~\ref{fig:PNomegabeta} the spectral power $\mathcal{P}_N^\nu (\omega)$ received by the last NP is shown for an SSH chain of 20 InSb particles where only the first particle is heated up to $310\,{\rm K}$ with respect to all other particles being at room temperature $300\,{\rm K}$. It can be nicely seen, that for the topological trivial case $\beta < 1$ the heat flux of the longitudinal modes is mainly due to the band modes in the lower frequency band, whereas the heat flux of the transversal modes is mainly due to the band modes in the upper frequency band. In the topological non-trivial case $\beta > 1$  the dominant contribution for $\mathcal{P}_N^{\nu}(\omega)$ is for $\nu = \parallel$ and $\nu = \perp$ given by the two edge modes which have nearly degenerate frequencies close to  $\omega_{\rm LP} = 1.752\times10^{14}\,{\rm rad}{\rm s}^{-1}$ and are therefore lying in the gap between the two bands. In Fig.~\ref{fig:PNomega} we show $\mathcal{P}_N(\omega)$ from Fig.~\ref{fig:PNomegabeta} for the two values $\beta = 0.7$ and $1.3$ separately to clearly demonstrate the band mode dominated heat flux in the trivial phase with $\beta = 0.7$ and the edge mode dominated heat flux in the non-trivial phase with $\beta = 1.3$, where for the two longitudinal edge modes $\omega_{\rm em,\parallel} \approx 1.7495\,{\rm rad/s}$ and for the two transversal edge modes $\omega_{\rm em,\perp} \approx 1.7493 \,{\rm rad/s}$. Note, that due to the large damping in the InSb NPs the modes are very broad and also broad band modes add to the edge mode dominated heat flux. By reducing the damping constant $\Gamma$ the contribution of the different modes would of course become much more narrowband which would allow for a better discrimation of band and edge modes.

\begin{figure}[hbt]
	\includegraphics[width=0.5\textwidth]{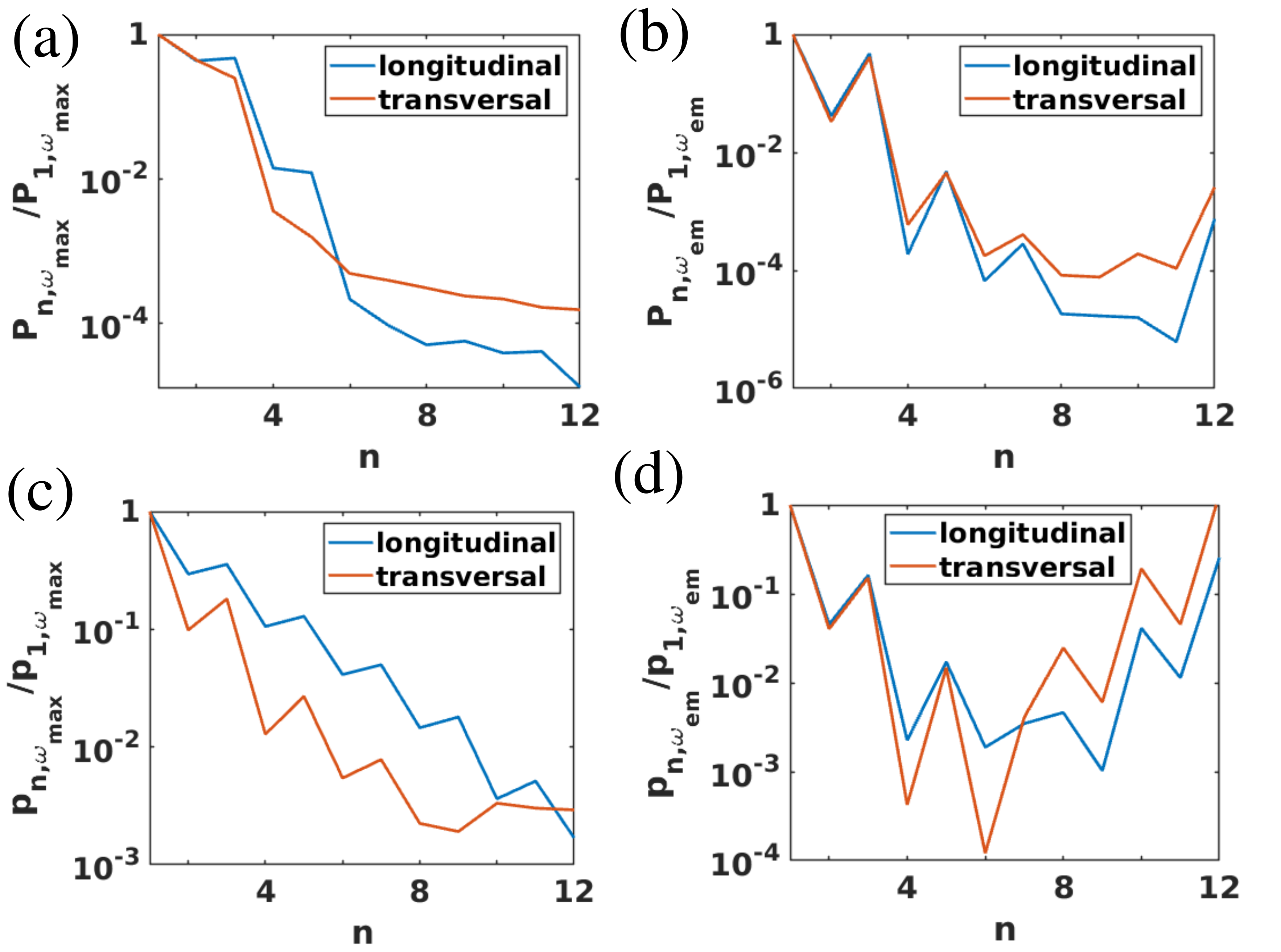}
	\caption{(a) $\mathcal{P}_n^\nu ( \omega_{\max} )$ at each NP position $n = 1,\ldots,12$ normalized to $\mathcal{P}_1^\nu (\omega_{\max})$ at the maximum frequency $\omega_{\max}$ in the band gap. (b) $\mathcal{P}_n^\nu (\omega_{\rm em})$ at each NP position $n = 1,\ldots,12$ normalized to $\mathcal{P}_1^\nu (\omega_{\rm rm})$ choosing a reduced damping $\Gamma = 10^{11}\,{\rm s}^{-1}$ at the edge mode frequency $\omega_{\rm em}$. (c) and (d) are the modulus of the real parts of the corresponding complex eigenvalues for the dipole moments to (a) and (b) of the SSH chain. $N = 12$, $R = 100\,{\rm nm}$, $d = 1\,\mu{\rm m}$, $\beta = 1.3$, and $T_1 = 310\,{\rm K}$,  and $T_{j} = 300\,{\rm K}$ ($j = 2, \ldots, 12$).
  }
  \label{fig:Dipoledistribution}
\end{figure}

Now, in order to make the impact of the edge modes on the heat flux along the NP chain more obvious, we show in Fig.~\ref{fig:Dipoledistribution} the power received by each NP along the SSH chain for a chain of $12$ NPs and $\beta = 1.3$. In Fig.~\ref{fig:Dipoledistribution}(a) we plot $\mathcal{P}_n^\nu (\omega_{\rm max})$ ($n = 1,\ldots,12$ at the frequency $\omega_{\rm max}$ where $\mathcal{P}_n^\nu (\omega)$ has its maximum within the band gap region.  In Fig.~\ref{fig:Dipoledistribution}(c) we show the corresponding eigenvalues of the dipole moments from Eq.~(\ref{Eq:Eigenmodes}) for the edge modes. In this dipole moment distribution a clear zig-zag behaviour can be seen which is due to the fact that there is a sublattice symmetry in the SSH model forbidding transitions between the lattices of the $A$ NPs and $B$ NPs. Due to the retardation and dissipation in our model, there is a small exitation of the $B$ NPs as well when the $A$ NPs are excited as can be nicely seen in Fig.~\ref{fig:Dipoledistribution}(c). The same trend as for the dipole moment distribution can also be seen in the distribution of the spectral power received by the different NPs in Fig.\ref{fig:Dipoledistribution}(a). This agreement between dipole moment distribution of the edge states and the spectral power distribution is even much clearer when the dissipation of the InSb particles is reduced. In Figs.~\ref{fig:Dipoledistribution}(b) and (d) we show these distributions for the edge modes by choosing a reduced damping $\Gamma  = 10^{11}\,{\rm s}^{-1}$. In these Figs.\ the coupling of the edge mode at the first particle $A$ and the last particle $B$ can be nicely seen. In particular the excitation of the last NP in the SSH chain by the edge modes is responsible for the large edge mode heat flux $\mathcal{P}_N^\nu$ received by the last NP in the SSH chain as observed in Fig.~\ref{fig:PNomegabeta}.

\begin{figure}[hbt]
	\includegraphics[width=0.45\textwidth]{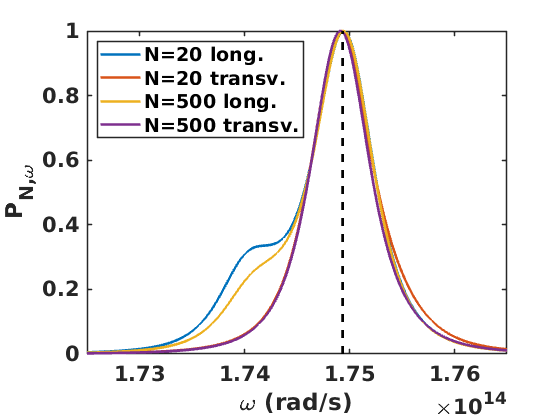}
	\caption{Normalized spectral power $\mathcal{P}_N^\nu ( \omega)$ for longitudinal and transversal modes for $N = 20$ and $N = 500$ with the edge mode resonances at $\omega_{\rm em,\perp} \approx \omega_{\rm em,\parallel} \approx 1.493\times10^{14}\,{\rm rad/s}$ (vertical dashed line). $R = 100\,{\rm nm}$, $d = 1\,\mu{\rm m}$, $\beta = 1.3$, and $T_1 = 310\,{\rm K}$,  and all other NP have temperatures $T_{j} = 300\,{\rm K}$ ($j = 2, \ldots, N$).
  }
  \label{fig:Chainlength}
\end{figure}

\begin{figure}[hbt]
	\includegraphics[width=0.45\textwidth]{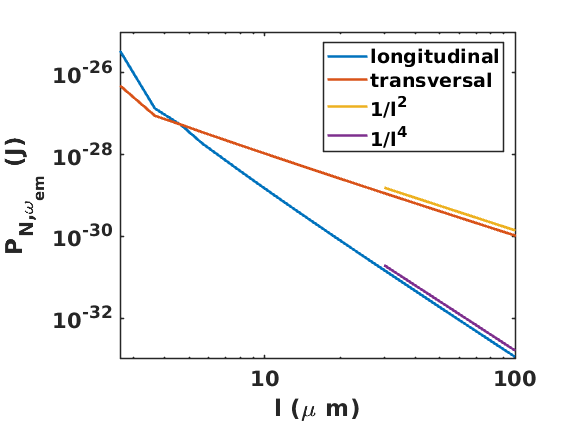}
	\caption{Spectral power $\mathcal{P}_N^\nu ( \omega_{\rm em})$ at the edge mode frequency $\omega_{\rm em}$ for longitudinal and transversal modes as function of the chain length $l$. Furthermore curves for a pure $1/l^2$ and $1/l^4$ dependence are included to show the power law scaling of $\mathcal{P}_N^\nu ( \omega_{\rm em})$. Here $R = 100\,{\rm nm}$, $d = 1\,\mu{\rm m}$, $\beta = 1.3$, and $T_1 = 310\,{\rm K}$,  and $T_{j} = 300\,{\rm K}$ ($j = 2, \ldots, N$).
  }
  \label{fig:ChainlengthDep}
\end{figure}

Finally, we want to study the dependence of the band and edge mode contribution to the heat flux on the length of the SSH chain. To this end, we plot in Fig.~\ref{fig:Chainlength} the spectral power $\mathcal{P}_N^\nu(\omega)$ received by the last particle normalized to the maximum value for $N = 20$ and $N = 500$ corresponding to two different chain lengths $l = d(N/2 - 1) + \beta d/2$. First of all, we note that the relative contribution of the band modes does not change much as function of the chain length. Obviously, in the topological phase the relative edge mode contribution at $\omega_{\rm em} = 1.743\times10^{14}\,{\rm rad/s}$ is fully invariant to the chain length and the normalized ``shape'' of the edge mode spectrum is the same for the longitudinal and transversal contribution. We find that the normalized profiles of $\mathcal{P}_N^\nu (\omega)$ are equivalent for all choice of $N$ as long as $N$ is larger than $8$. That means we can conclude that the chain length dependence of the band and edge modes is the same. In Fig.~\ref{fig:ChainlengthDep} it can be seen that the heat flux for longitudinal edge modes scale like $1/l^4$ whereas for the transversal edge modes it scale like $1/l^2$. We find the same scaling laws for the band modes. Furthermore, we can see from Fig.~\ref{fig:Chainlength} that even for chains with $N = 500$ which corresponds to a total length of about $250 \,\mu {\rm m} \approx 25 \lambda_{\rm th}$ the edge modes dominate the radiative heat transfer. This is a peculiar feature, because naively one would expect that the edge modes cannot contribute for long chaines. On the other hand, since the edge mode and band mode contributions obey the same scaling law for the heat flux with respect to the chain length, we can conclude that both heat flux channels provided by band and edge modes are equivalent for the radiative heat flux.  Hence, even though the edge modes are confined to the edges of the chain as can be seen in Fig.~\ref{fig:Dipoledistribution}(d) the heat flux through the chain by the edge mode coupling is giving the major heat flux channel for any chain length.

To understand the nature of the coupling, it suffices to realize that for both the edge and band modes the coupling is dictated by the dipole-dipole coupling in $G_\perp(x_N - x_1)$ for the transversal and $G_\parallel(x_N - x_1)$ for the longitudinal modes. For $k_0 |x_N - x_1| = k_0 l \gg 1$ which is easily fulfilled for $d = 1\,\mu{\rm m}$ and chains with more than 20 particles and $\lambda \approx \lambda_{\rm th}$ we are in the retarded regime and have $G_\perp(x_N - x_1) \propto 1/l$ and $G_\perp(x_N - x_1) \propto 1/l^2$. The transmission coefficient $\mathcal{T}_{1 \rightarrow N}, \nu \propto |G_\nu (x_N - x_1)|^2$ in Eq.~(\ref{Eq:TransmissionKoeff}) connecting the first and the last NP is then scaling like $1/l^2$ for the transversal and $1/l^4$ for the longitudinal modes no matter if the modes are band or edge modes. Hence, the fact that the edge modes contribute for the heat flux throug a long chain can be traced back to the retarded dipole-dipole coupling of the first and last particle.


\section{Conclusion}

We have discussed the radiative heat flux along an SSH chain of InSb NPs. We have found that in the topological phase the radiative heat flux along a SSH NP chain can be dominated by the contribution of the edge modes. This is an unexpected effect because the edge modes at both ends of the chain are not allowed to propagate. One can argue that they can still couple via their evanescent fields and in analogy to coupled surface modes they can contribute to the radiative heat flux. However, we find that despite the strong confinement of the edge modes there is a long-range coupling between the first and the last NP in the SSH chain due to the inclusion of retardation in our model which is the reason for the edge mode dominated heat flux is for any chain length. This peculiar effect of the dominance of the edge modes shows that the topological edge modes are not excluded from radiative heat transfer, but on the contrary can serve as an important heat flux channel along an SSH chain in its topological phase. Our work lays the foundation for future investigations of near-field radiative heat transfer in topological systems.

\acknowledgments
\noindent

S.-A.\ B.\ thanks Philippe Ben-Abdallah for inspiring discussions on ``topological heat radiation'' over the last five years and acknowledges support from Heisenberg Programme of the Deutsche Forschungsgemeinschaft (DFG, German Research Foundation) under the project No. 404073166.


\begin{thebibliography}{99}
   \bibitem{PBAmanybody} P. Ben-Abdallah, S.-A. Biehs, and K. Joulain, Phys. Rev. Lett. {\bf 107}, 114301 (2011).
  \bibitem{KruegerEtAl2012} M. Kr\"{u}ger, G. Bimonte, T. Emig \and M. Kardar, Phys. Rev. B {\bf 86}, 115423 (2012).
  \bibitem{nteilchen} R. Messina. M. Tschikin, S.-A. Biehs and P. Ben-Abdallah, Phys. Rev. B, \textbf{88},104307 (2013).
  \bibitem{DongEtAl2017} J. Dong, J.Zhao, L. Liu, Phys. Rev. B {\bf 95}, 125411 (2017).
  \bibitem{LatellaEtAl2017}  I. Latella, P. Ben-Abdallah, S.-A. Biehs, M. Antezza, and R. Messina, Phys. Rev. B {\bf 95}, 205404 (2017). 
  \bibitem{PBAEtAl2013} P. Ben-Abdallah, R. Messina, S.-A. Biehs, M. Tschikin, K. Joulain, and C. Henkel, Phys. Rev. Lett. {\bf 111}, 174301 (2013).
  \bibitem{LatellaEtAl2018} I. Latella S.-A. Biehs, R. Messina, A. W. Rodriguez, and P. Ben-Abdallah, Phys. Rev. B {\bf 97}, 035423 (2018).
  \bibitem{Kathmann} C. Kathmann, R. Messina, P. Ben-Abdallah, S.-A. Biehs, Phys. Rev. B {\bf 98}, 115434 (2018).
  \bibitem{Tervo} E. Tervo, M. Francoeur, B. A. Cola, and Z. M. Zhang, Phys. Rev. B {\bf 100}, 205422 (2019). 
  \bibitem{Tervo2} E. J. Tervo and B. A. Cola and Z. M. Zhang, JQSRT {\bf 246}, 106947 (2020).
  \bibitem{Drag} P. Ben-Abdallah, Phys. Rev. B {\bf 99}, 201406 (2019).
  \bibitem{Add4}R. Incardone, T. Emig, and M. Kr\"{u}ger, Europhys. Lett. \textbf{106}, 41001 (2014).
  \bibitem{Add5}M. Nikbakht, Europhys. Lett. \textbf{110}, 14004 (2015).
\bibitem{PBASAB2014} P. Ben-Abdallah and S.-A. Biehs, Phys. Rev. Lett. {\bf 112}, 044301 (2014).
\bibitem{OrdonezEtAl2016} J. Ordonez-Miranda, Y. Ezzahri, J. Drevillon, and K. Joulain, Phys. Rev. Appl. {\bf 6}, 054003 (2016).
\bibitem{KathmannEtAl2020} C. Kathmann, M. Reina, R. Messina, P. Ben-Abdallah, and S.-A. Biehs, Sci. Rep. {\bf 10}, 3596 (2020).
  \bibitem{Carlos} J. C. Cuevas and F. J. Garc{\'\i}a-Vidal, ACS Photonics {\bf 5}, 3896 (2018).
  \bibitem{RMP} S.-A. Biehs, R. Messina, P. S. Venkataram, A. W. Rodriguez, C. Cuevas, P. Ben-Abdallah, arXiv:2007.05604. 
\bibitem{Edalatpour2014} S. Edalatpour and M. Francoeur, J. Quant. Spectr. Radiat. Transfer {\bf 133}, 364 (2014).
\bibitem{EkerothEtAl2017} R.~M.~A. Ekeroth, A.  Garc\'{i}a-Martin, and J.-C. Cuevas, Phys. Rev. B {\bf 95}, 235428 (2017).
\bibitem{moncadavilla} E. Moncada-Villa, V. Fernández-Hurtado, F. J. Garcia-Vidal,  A. Garc\'{i}a-Mart\'{i}n and J.C. Cuevas, Phys. Rev. B \textbf{92}, 125418 (2015). 
  \bibitem{Song} J. Song and Q. Cheng, Phys. Rev. B {\bf 94}, 125419 (2016).
  \bibitem{WuEtAl} H. Wu, Y. Huang, L. Cui, K. Zhu, Phys. Rev. Appl. {\bf 11}, 054020 (2019).
  \bibitem{Herz} F. Herz, S.-A. Biehs, EPL {\bf 127}, 4 (2019).
  \bibitem{zhufan} L. Zhu and S. Fan, Phys. Rev. Lett. {\bf 117}, 134303 (2016).
  \bibitem{zhufan2} L. Zhu and S. Fan, Phys. Rev. B { \bf 97}, 094302 (2018).
  \bibitem{Silveirinha} M. G. Silveirinha, Phys. Rev. B {\bf 95}, 115103 (2017).
  \bibitem{meinpaper} A. Ott, P. Ben-Abdallah, and S.-A. Biehs, Phys. Rev. B {\bf 97}, 205414 (2018).
  \bibitem{Zubin2019} C. Khandekar, Z. Jacob, New J. Phys. {\bf 21}, 103030 (2019).
  \bibitem{Latella2017} I. Latella and P. Ben-Abdallah, Phys. Rev. Lett. {\bf 118}, 173902, (2017).
  \bibitem{Cuevas} R. M. Abraham Ekeroth, P. Ben-Abdallah, J.C. Cuevas, and A. Garcia Martin, ACS Photonics {\bf 5}, 705 (2017).
  \bibitem{hall} P. Ben-Abdallah, Phys. Rev. Lett. {\bf 116}, 084301, (2016).
  \bibitem{OttEtAl2019} A. Ott, R. Messina, P. Ben-Abdallah, and S.-A. Biehs, J. Photon. Energy {\bf 9}, 032711 (2019).
  \bibitem{OttEtAl2020} A. Ott, S.-A. Biehs, and P. Ben-Abdallah, Phys. Rev. B {\bf 101}, 241411(R) (2020).
  \bibitem{paper_diode} A. Ott, R. Messina, P. Ben-Abdallah and S.-A. Biehs, Appl. Phys. Lett. \textbf{114}, 163105 (2019).
  \bibitem{Fiorino2018} A. Fiorino, D. Thompson, L. Zhu, B. Song, P. Reddy and E. Meyhofer, Nano Lett. {\bf 18}, 3711 (2018).
  \bibitem{OESSH} C. W. Ling, M. Xiao, C. T. Chan, S. F. Yu, and K. H. Fung, Opt. Expr. {\bf 23}, 225887 (2015)
  \bibitem{ACSphotonSSH} S. R. Pocock, X. Xiao, P. A. Huidobro, and V. Giannini, ACS Photonics {\bf 5}, 2271 (2018).
  \bibitem{JAPSSH} B. X. Wang and C. Y. Zhao, J. Appl. Phys. {\bf 127}, 073106 (2020).
  \bibitem{Novotny} L. Novotny and B. Hecht, {\em Principles of Nano-Optics}, Cambridge University Press, (2006).
  \bibitem{ford} W.~H. Weber and G.~W. Ford, Phys. Rev. B \textbf{70}, 125429 (2004).
  \bibitem{exp} S. Law, R. Liu, and D. Wasserman, J. Vac. Sci. Technol. B  \textbf{32}, 052601 (2014).
  \bibitem{Naraynaswamy2008} A. Narayanaswamy and G. Chen, Phys. Rev. B {\bf 77}, 075125 (2008).
  \bibitem{PBA2008} P. Ben-Abdallah, K. Joulain, J. Drevillon, and C. Le Goff, Phys. Rev. B {\bf 77},  075417 (2008).
  \bibitem{Otey} C. Otey, S. Fan, Phys. Rev. B {\bf 84}, 245431 (2011).
  \bibitem{Becerril} D. Becerril, C. Noguez, Phys. Rev. B {\bf 99}, 045418 (2019).
\end{thebibliography}
\end{document}